\documentclass[a4paper,12pt]{article}
\usepackage{epsfig,amssymb,amsmath}
\usepackage{graphicx}
\usepackage{amsfonts}

\usepackage{tikz}
\usepackage{tkz-euclide}
\tikzset{circ/.style = {fill, circle, inner sep = 0, minimum size = 3}}

\tikzset{eqpic/.style={baseline={([yshift=-.5ex]current bounding box.center)}}}

\definecolor{mblue}{rgb}{0.2, 0.3, 0.8}
\begin{document}

\newcommand{\be}{\begin{equation}}
\newcommand{\ee}{\end{equation}}
\newcommand{\bea}{\begin{eqnarray}}
\newcommand{\eea}{\end{eqnarray}}
\newcommand{\bean}{\begin{eqnarray*}}
\newcommand{\eean}{\end{eqnarray*}}
\font\upright=cmu10 scaled\magstep1
\font\sans=cmss12
\newcommand{\ssf}{\sans}
\newcommand{\stroke}{\vrule height8pt width0.4pt depth-0.1pt}
\newcommand{\Z}{\hbox{\upright\rlap{\ssf Z}\kern 2.7pt {\ssf Z}}}
\newcommand{\ZZ}{\Z\hskip -10pt \Z_2}
\newcommand{\C}{{\rlap{\upright\rlap{C}\kern 3.8pt\stroke}\phantom{C}}}
\newcommand{\R}{\hbox{\upright\rlap{I}\kern 1.7pt R}}
\newcommand{\HH}{\hbox{\upright\rlap{I}\kern 1.7pt H}}
\newcommand{\CP}{\hbox{\C{\upright\rlap{I}\kern 1.5pt P}}}
\newcommand{\identity}{{\upright\rlap{1}\kern 2.0pt 1}}
\newcommand{\half}{\frac{1}{2}}
\newcommand{\quart}{\frac{1}{4}}
\newcommand{\pr}{\partial}
\newcommand{\bm}{\boldmath}
\newcommand{\I}{{\cal I}} 
\newcommand{\M}{{\cal M}}
\newcommand{\N}{{\cal N}}
\newcommand{\e}{\varepsilon}
\newcommand{\x}{{\bf x}}
\newcommand{\lam}{\lambda}
\newcommand{\nn}{\nonumber}

\thispagestyle{empty}
\vskip 3em
\begin{center}
{{\bf \huge Robustness of the Hedgehog Skyrmion}} 
\\[15mm]

{\bf \Large N.~S. Manton\footnote{email: N.S.Manton@damtp.cam.ac.uk;
    ORCID: 0000-0002-2938-156X}} \\[20pt]

\vskip 1em
{\it 
Department of Applied Mathematics and Theoretical Physics,\\
University of Cambridge, \\
Wilberforce Road, Cambridge CB3 0WA, U.K.}
\vspace{12mm}

%%%%%%%%%%%%%%%%%%%%%%%%%%%%%%%%%%%%%%%%%%%%%%%%%%
\abstract
{We investigate the radial profile function of the hedgehog Skyrmion with 
unit baryon number in generic EFTs (effective field theories) of pions. The
analysis assumes chiral symmetry and ignores the pion mass term. The
Skyrmion is always smooth, because it has no point
source at the origin, and terms in the EFT with higher numbers of pion
derivatives do not result in uncontrolled large corrections or
singularities there. The profile varies in quite a limited way as the
terms in the EFT change, and a universal profile function is proposed. 
}
%%%%%%%%%%%%%%%%%%%%%%%%%%%%%%%%%%%%%%%%%%%%%%%%%%

\end{center}

\vskip 150pt
\leftline{Keywords: Hedgehog Skyrmion, Radial profile, Effective field
theory}
\vskip 1em

\vfill
\newpage
\setcounter{page}{1}
\renewcommand{\thefootnote}{\arabic{footnote}}

%%%%%%%%%%%%%%%%%%%%%%%%%%%%%%%%%%%%%%%%%%%%%%%%%%%%%%%%%%%%%%%%%%%%%%%%%%%%%
%%%%%%%%%%%%%%%%%%%%%%%%%%%%%%%%%%%%%%%%%%%%%%%%%%%%%%%%%%%%%%%%%%%%%%%%%%%%%

\section{Introduction} 
\vspace{2mm}

In the 1960s, Skyrme introduced a nonlinear field theory of pions whose
static classical solutions are topological solitons -- Skyrmions
\cite{Sk}. Skyrmions are reviewed in the books \cite{Weig,Shn,Man}.
A Skyrmion's topological charge is identified with its baryon
number (atomic mass number, or nucleon number) $B$, and the
fundamental solution has $B=1$. This $B=1$ solution's fields are spherically
symmetric, and because the triplet of pion fields points radially
outwards, it is often referred to as a ``hedgehog''. A
hedgehog configuration is completely specified by a radial
profile function $f(r)$ satisfying the boundary conditions $f(0) = \pi$
and $f(\infty) = 0$. The static field equations reduce to a second-order
ODE for $f$, which has a unique solution satisfying these boundary
conditions. The hedgehog Skyrmion, rather than a configuration with
less symmetry, is almost certainly the minimal-energy static solution
in the $B=1$ sector of the field theory. When its rotational degrees
of freedom are quantized, one finds low-energy states representing
spin-half protons and neutrons \cite{ANW}.

Skyrme's original Lagrangian has three terms, a term quadratic in first
derivatives of the field (the Dirichlet term), a term quartic in first
derivatives (the Skyrme term), and a pion mass term with
no derivatives. When the pion mass term is dropped, the theory has full
$SU(2)_{\rm L} \times SU(2)_{\rm R}$ chiral symmetry, although this is
spontaneously broken to $SO(3)$ isospin symmetry by the vacuum field
configuration. It is easier to analyse the Skyrmion's profile $f(r)$  
for massless pions, where the large-$r$ tail has power rather
than exponential decay, and this is what we do throughout this paper.

The balance between terms quartic and quadratic in derivatives gives the
Skyrmion a finite size in three spatial dimensions.
Furthermore, the particular term quartic in derivatives chosen by
Skyrme has the property that in the Lagrangian,
time-derivatives only appear quadratically. Subsequently, it was
realised that there is one further term, sextic in derivatives, that
is compatible with chiral symmetry and is still only quadratic in
time-derivatives \cite{JJGBC}. This term is often included in
Skyrme-type Lagrangians. It is called the BPS-Skyrme term.

In parallel to studies of Skyrmions, chiral effective field theories
(EFTs) of pions and nucleons have been developed using perturbation theory
\cite{Wei,Leu,ORvK,EM}. For a recent review and comprehensive
bibliography, see \cite{Epe}. In the perturbative picture, nucleons have
their own fundamental fields which act as sources for the pion
fields, and the configurations have
no topological stability. The terms in the EFT Lagrangian are not
constrained to have only two time-derivatives, and several more terms
are included. Their coefficients are fitted to data extracted (mainly) from
nucleon-nucleon scattering. As the (mass) dimensions of terms
increase with the number of derivatives, the coefficients
incorporate increasing powers of the inverse of a mass cut-off.
Terms beyond a certain number of derivatives can therefore be neglected
provided one is primarily interested in physics at momenta and
energies below this cut-off scale.

From the general EFT perspective, the analysis of Skyrme's original field
equations has been criticised on the grounds that many terms that could be
present in the Lagrangian have been ignored. Moreover, the balance
between the quartic and quadratic terms, contributing
to the Skyrmion's energy, suggests that it is
wrong to neglect terms with higher powers of derivatives, and that
these terms could substantially change the Skyrmion's core structure.
This criticism relies on the idea that the spatial derivatives of the
fields are large in the core region, close to the nucleon sources at $r=0$.
However, the $B=1$ Skyrmion in Skyrme's
original theory is quite smooth at $r=0$ because there is no point
source. So a more reasonable expectation is that variations in the
coefficients of the EFT will make small changes to the Skyrmion profile. This
is what we aim to show here.

In detail, we will consider the most general terms that arise in a
pion EFT with chiral symmetry, and their combination into the energy
functional for a static, hedgehog $B=1$ Skyrmion. We will allow arbitrary
powers of the first radial derivative of the profile, $f'(r)$, but exclude
$f''(r)$ and higher derivatives of $f$ from the energy density. To avoid
excessive algebra, we will mostly ignore terms that are more than sextic in
$f'$, but show that it would be straightforward to include them. Some of the
terms we retain, quartic or sextic in $f'$, occur in perturbative EFT,
but have previously been excluded from most analyses of Skyrmions
because they are partnered by terms in the full Lagrangian that are
quartic or sextic in time-derivatives. We do not discuss the
potentially complicated time-evolution and quantization that is
associated with such time-derivatives. Their effects are probably
controllable if the time-derivatives are small.

We will solve the second-order nonlinear ODE for $f$ up to $O(r^3)$
for small $r$, and up to $O\left(\frac{1}{r^6}\right)$ for large $r$.
These approximate solutions can be matched at some specific, adjustable
radius $r=R$. The result is a continuous profile with continuous slope.
For a special choice of the coefficient of the $O(r^3)$ term for small
$r$, generally different
from the coefficient arising in a particular EFT, $f$ has a continuous
second derivative, i.e. continuous curvature. Aside from the spatial
scale $R$, this is a ``universal'' profile function that can be used as an
approximate ansatz for a hedgehog Skyrmion in just about any EFT.
Local analysis does not fix $R$; it needs to be determined by
minimising the total energy in an EFT of interest. This requires
some numerical integration, which we perform in the simplest case.

In our analysis, we fix the Dirichlet and Skyrme terms to have
standard, conventional coefficients. However, physically, these
coefficients should be calibrated to fit the size and mass of a
nucleon. So, although the scale $R$ varies between different
Skyrme-type theories, there is freedom to calibrate
all Skyrmions to have the same leading large-$r$ behaviour. Only the
detailed shape of the core profile depends on the variant, and this
shape does not vary much. In this sense, the hedgehog Skyrmion is
robust. Further numerical work would be needed to verify this claim
more completely. 

Our whole approach has a lot in common with that of Marleau in
ref.\cite{Mar}. Marleau also solved for the hedgehog profile $f(r)$
at small $r$ and large $r$, generating recursion relations for the
coefficients of the powers of $r$ and $\frac{1}{r}$ respectively, to
all orders, and wrote down a very general profile that matches both
asymptotic expansions. However, the resulting profile still
depends on three undetermined parameters, a slope at $r=0$, a
leading coefficient at large $r$, and a scale $R$. These three parameters
need to be calculated numerically, by energy minimisation. This
contrasts with our analysis, where matching is used to fix two of
these parameters, at the price of $f$ not being smooth to all orders.
An important further difference is that Marleau's energy functional
is restricted to being quadratic in $f'$. By contrast, we are
allowing chirally-invariant terms including higher powers of $f'$,
that arise from more general EFTs. Our classification of
chirally-invariant energy densities is essentially the same as that
of Gudnason and Nitta \cite{GN}, but they did not simultaneously
consider the $B=1$ hedgehog profile. 

\section{Energy Functionals}
\vspace{2mm}

In this section, we exploit the formalism of the strain tensor of a
Skyrme field to construct general Skyrme energy functionals for a static $B=1$
hedgehog configuration with profile $f(r)$ \cite{Ma2}.

Recall that a (static) Skyrme field $U(\x)$ is an $SU(2)$-valued field that can
be expressed in terms of triplet pion and singlet sigma fields as
\be
U(\x) = \sigma({\x}) {\boldsymbol{1}}_2 + i {\boldsymbol{\pi}}({\x}) \cdot
{\boldsymbol{\tau}} \,,
\ee
with ${\boldsymbol{1}}_2$ the unit $2 \times 2$ matrix and
${\boldsymbol{\tau}}$ the Pauli matrices. The constraint
\be
\sigma^2 + {\boldsymbol{\pi}}\cdot{\boldsymbol{\pi}} = 1
\ee
also needs to be satisfied (since $SU(2)$, as a manifold, is a 3-sphere),
so only the pion fields ${\boldsymbol{\pi}}$ are independent.
The current $R_i = \pr_iU U^{-1}$ plays the role of the Skyrme field
gradient, and a key quantity is the strain tensor $D_{ij} = -\half
{\rm Tr}(R_iR_j)$, which is a positive-definite geometrical measure
of the Skyrme field's spatial variation. It characterises the local
elastic strain of the mapping $U$ from spatial $\R^3$ to the unit
3-sphere of $SU(2)$, equipped with their standard Riemannian metrics.
The strain tensor is invariant under global chiral transformations
$U({\x}) \to U_{\rm L} U({\x}) U_{\rm R}^{-1}$. The
eigenvalues of the strain tensor are also invariant under local
spatial rotations, and symmetric polynomials in these eigenvalues are
candidates for the Skyrme field's energy density.

Let us denote these eigenvalues by $\lam_1^2, \lam_2^2, \lam_3^2$. The Skyrme
energy functional is parity-invariant, so it involves only even, symmetric
polynomials in $\lam_1, \lam_2, \lam_3$. A fundamental algebraic
basis for such polynomials is
\bea
I_2 &=& \lam_1^2 + \lam_2^2 + \lam_3^2 \,, \nn \\
I_4 &=& \lam_1^4 + \lam_2^4 + \lam_3^4 \,, \nn \\
I_6 &=& \lam_1^6 + \lam_2^6 + \lam_3^6 \,.
\label{polyI}
\eea
Also useful are, alternatively,
\bea
J_2 = I_2 &=& \lam_1^2 + \lam_2^2 + \lam_3^2 \,, \nn \\
J_4 = \half (I_2^2 - I_4) &=& \lam_1^2\lam_2^2 + \lam_2^2\lam_3^2
+ \lam_3^2\lam_1^2 \,, \nn \\
J_6 = \frac{1}{6} (I_2^3 - 3I_2I_4 + 2I_6) &=& \lam_1^2\lam_2^2\lam_3^2 \,.
\label{polyJ}
\eea

The hedgehog ansatz for the Skyrme field is
\be
U_{\rm H}({\x}) = \exp \left(if(r) \, \hat\x \cdot {\boldsymbol{\tau}}
\right) = \cos f(r) {\boldsymbol{1}}_2 + i\sin f(r) \, \hat\x
\cdot {\boldsymbol{\tau}} \,,
\ee
where $\hat\x$ is the spatial, outward radial unit-vector. $f(r)$ is the
radial profile function, required to satisfy the boundary
conditions $f(0) = \pi$ and $f(\infty) = 0$, so that
$U_{\rm H} = -{\boldsymbol{1}}_2$ at the origin
and $U_{\rm H} = {\boldsymbol{1}}_2$ (the vacuum value) at spatial
infinity. For the hedgehog ansatz, the strain tensor has eigenvalues
\be
\lam_1^2 = f'^2 \,, \quad \lam_2^2 = \lam_3^2 = \frac{\sin^2 f}{r^2}
\label{hhogevals}
\ee
(where $f$ denotes $f(r)$ from here on).
The first is the strain in the radial direction, and the second
and third are the two equal strains in the angular directions. $\sin
f$ rather than $f$ occurs here because of the curvature of the target
$SU(2)$. The fundamental symmetric polynomials (\ref{polyI}) and
(\ref{polyJ}) reduce to
\bea
I_2 &=& f'^2 + 2 \, \frac{\sin^2 f}{r^2} \,, \nn\\
I_4 &=& f'^4 + 2 \, \frac{\sin^4 f}{r^4} \,, \nn \\
I_6 &=& f'^6 + 2 \, \frac{\sin^6 f}{r^6} \,,
\label{Iformula}
\eea
and
\bea
J_2 &=& f'^2 + 2 \, \frac{\sin^2 f}{r^2} \,, \nn \\
J_4 &=& 2 f'^2 \, \frac{\sin^2 f}{r^2} + \frac{\sin^4 f}{r^4} \,, \nn \\
J_6 &=& f'^2 \, \frac{\sin^4 f}{r^4} \,.
\label{Jformula}
\eea
The polynomials $J_2, J_4, J_6$ are quadratic in $f'$, so the
corresponding terms in the full Lagrangian are quadratic in
time-derivatives. It is therefore a linear combination of these
polynomials that gives the traditional energy density for a Skyrme-type
theory, and Skyrme's original theory combines just $J_2$ (the Dirichlet
term) and $J_4$ (the Skyrme term). $J_6$ is the additional (BPS-Skyrme) term
introduced by Jackson et al. \cite{JJGBC}. Here we will allow for terms
with higher powers of these polynomials, and therefore higher powers
of $f'$. It is then often more convenient to use the simpler monomials
\be
K_{k,l} = f'^k \frac{\sin^l f}{r^l} \,, 
\ee
where $k$ and $l$ are even integers (not both
zero). However, individually $K_{k,l}$ is not a symmetric
polynomial in the strain eigenvalues, so is not chirally invariant.
Symmetrisation is required.

Our first result concerns the allowed behaviour of the profile $f$ near
$r=0$. We show, quite generally, that $f$ has leading dependence
\be
f(r) = \pi - \beta r + O(r^3) \,,
\label{fsmall}
\ee
with $\beta$ undetermined locally. This result holds provided that the
energy functional is constructed from the symmetric polynomials
$I_2, I_4, I_6$, but does not hold for arbitrary combinations of the monomials
$K_{k,l}$. Consider a general polynomial energy density of the form
$G(I_2,I_4,I_6)$. For the hedgehog ansatz, this density is independent
of angles and only depends on $r$. The total energy is therefore
\be
E = 4\pi \int_0^\infty r^2 G(I_2,I_4,I_6) \, dr \,.
\ee
$I_2,I_4,I_6$ depend on $f'$ and $\sin f$, as in
formulae (\ref{Iformula}). Therefore the
Euler--Lagrange equation for $f$ has the form
\bea
&& \frac{d}{dr}\left\{ r^2 \left(
2\frac{\pr G}{\pr I_2} f' + 4\frac{\pr G}{\pr I_4} f'^3 +  
6\frac{\pr G}{\pr I_6} f'^5 \right) \right\} \nn \\
&& \quad - 2r \left( 2\frac{\pr G}{\pr I_2}\frac{\sin f}{r} 
+ 4\frac{\pr G}{\pr I_4}\frac{\sin^3 f}{r^3} 
+ 6\frac{\pr G}{\pr I_6}\frac{\sin^5 f}{r^5} \right) \cos f = 0 \,.
\label{geneq}
\eea
This equation is invariant under the transformations $r \to -r$ and $f
\to f + 2\pi$, so it is consistent for $\pi - f(r)$ to be a
series in odd powers of $r$ for small $r$, with no term proportional
to $r^2$. Moreover, eq.(\ref{fsmall}) implies that
$f' = -\beta + O(r^2)$, $\frac{\sin f}{r} = \beta + O(r^2)$ and $\cos f
= -1 + O(r^2)$, so $I_2, I_4, I_6$ and $f'$ are constants with $O(r^2)$
corrections for small $r$, and have vanishing derivatives at $r=0$.
To linear order in $r$, it is therefore sufficient in eq.(\ref{geneq})
to differentiate $r^2$, obtaining $2r$; the equation is then clearly
satisfied at linear order, for any $G$, because $f' = -\beta$,
$\frac{\sin f}{r} = \beta$ and $\cos f = -1$ at $r=0$.

All the strain eigenvalues at $r=0$ are $\beta^2$, so a hedgehog profile
has an isotropic strain tensor there, i.e. the Skyrme field $U$ is
smooth and is locally a conformal map from physical $\R^3$ to the
target $SU(2)$. Note that the symmetrisation with respect to the
strain eigenvalues is crucial for this. For example, the energy functional
\be
\tilde E = 4\pi \int_0^\infty r^2 f'^2 \, dr
\ee
has Euler--Lagrange equation $\frac{d}{dr} (r^2 f') = 0$ and this is
not satisfied at $O(r)$ by $f$ of the form (\ref{fsmall}) if $\beta
\ne 0$. 

\section{Standard Hedgehog Skyrmion Revisited}
\vspace{2mm}

Here, we review the hedgehog Skyrmion profile $f(r)$ in the original
Skyrme theory with its Dirichlet and Skyrme terms, and massless
pions. This is an opportunity to develop our approach, and clarify
the approximations that it entails. The energy functional, expressed
in terms of the strain eigenvalues, is
\be
E = \int_{\R^3} (J_2 + J_4) \, d^3x
= \int_{\R^3} \left( \lam_1^2 + \lam_2^2 + \lam_3^2 +
\lam_1^2\lam_2^2 + \lam_2^2\lam_3^2 + \lam_3^2\lam_1^2 \right) \, d^3x \,.
\ee
For the hedgehog ansatz, with strain eigenvalues (\ref{hhogevals}), this
simplifies to
\be
E = 4\pi \int_0^\infty \left( r^2 f'^2 + 2\sin^2 f
  + 2f'^2 \sin^2 f + \frac{\sin^4 f}{r^2} \right) \, dr \,.
\label{SkhhogE}
\ee
The resulting field equation is
\be
r^2 f'' + 2r f' - \sin 2f
+ 2\sin^2 f \, f'' + \sin 2f \, f'^2 - \sin 2f \, \frac{\sin^2 f}{r^2} = 0 \,. 
\label{SkhhogEq}
\ee

For large $r$, a consistent local solution satisfying the
boundary condition $f(\infty) = 0$ is a power
series in $\frac{1}{r^2}$. The leading terms are of the form
\be
f(r) = \frac{C}{r^2} + \frac{D}{r^6} + O\left( \frac{1}{r^8} \right) \,,
\ee
and substituting this into (\ref{SkhhogEq}), one finds that
$C$ is arbitrary and $D = -\frac{1}{21} C^3$, so
\be
f(r) = \frac{C}{r^2} - \frac{1}{21} \frac{C^3}{r^6}
+ O\left( \frac{1}{r^8} \right) \,.
\ee
It is the $\sin f$ factor in the angular strain eigenvalues that
generates the $O(\frac{1}{r^6})$ correction to the leading $\frac{C}{r^2}$
term. The absence of a term proportional to $\frac{1}{r^4}$ is
because the series for $\sin f$ contains only odd powers of $f$.  

The most useful observation is that it is only the Dirichlet term,
contributing the first three terms to eq.(\ref{SkhhogEq}), and
not the Skyrme term, that determines the large-$r$ solution up to
$O(\frac{1}{r^6})$. The Skyrme term generates sources that affect
the coefficient of the $O(\frac{1}{r^8})$ term. This is a general
result that we will verify below -- no terms in the energy functional
apart from the Dirichlet term affect the large-$r$ behaviour of $f(r)$ up
to $O(\frac{1}{r^6})$. In this sense, we have established the long-range
robustness, or universality, of the hedgehog Skyrmion, as any
$O(\frac{1}{r^8})$ correction to a leading $O(\frac{1}{r^2})$
term is effectively a short-range correction. We will therefore,
from now on, always approximate the hedgehog profile for large $r$ by
\be
f(r) = \frac{C}{r^2} - \frac{1}{21} \frac{C^3}{r^6} \,.
\label{ftail}
\ee
This should be a good approximation in the range
$\frac{C}{r^2} \lesssim 2$, i.e. $r^2 \gtrsim \half C$. The
coefficient $C$ is ultimately fixed so that $f(0) = \pi$; in this sense,
$C$ does depend on all the terms in the energy functional.

Next, we consider the form of $f(r)$ for small $r$. Taking account of
the boundary condition $f(0) = \pi$, we write $f = \pi - g$, with
$g(0) = 0$. Then $f' = -g'$, $\sin f = \sin g$ and $\cos f = -\cos g$, so
the field equation for $g$ is
\be
r^2 g'' + 2r g' - \sin 2g + 2\sin^2 g \, g''
+ \sin 2g \, g'^2 - \sin 2g \, \frac{\sin^2 g}{r^2} = 0 \,, 
\label{SkhhoggEq}
\ee
the same as eq.(\ref{SkhhogEq}) for $f$. A consistent
solution is a series in odd powers of $r$, conveniently written as
\be
g(r) = \beta r - \gamma \beta^3 r^3 + O(r^5) \,.
\label{gcubic}
\ee
Substituting this into eq.(\ref{SkhhoggEq}), and working up to $O(r^3)$ gives
\be
\left(20\gamma - \frac{8}{3} + 40\gamma\beta^2
- \frac{4}{3}\beta^2 \right) \beta^3 r^3 = 0 \,.
\label{Skgammabeta}
\ee
As anticipated from the general argument of the last section, all the
terms linear in $r$ (with coefficients both $\beta$ and $\beta^3$)
cancel, so there is no local constraint on $\beta$. Equation
(\ref{Skgammabeta}) requires that
\be
\gamma = \frac{2}{15} \left(\frac{1+\half\beta^2}{1+2\beta^2}\right)
\label{Skgamma}
\ee
so
\be
f(r) = \pi -\beta r
+\frac{2}{15} \left(\frac{1+\half\beta^2}{1+2\beta^2}\right)
\beta^3 r^3 + O(r^5) \,.
\label{fcubic}
\ee
$\beta$ is ultimately determined by the need to satisfy the boundary condition
$f(\infty)=0$.

Even in the absence of the Skyrme term contributions, there is still
a local solution $f(r)$, with the bracketed expression in (\ref{fcubic})
replaced by $1$. This plays a role in the bag-like Skyrmions
of Gustafsson and Riska \cite{GR}. For more general energy functionals,
we shall find that $\gamma$ is always $\frac{2}{15}$ times a ratio of
polynomials in $\beta^2$ with leading coefficients $1$. 

\section{Further Energy Contributions}
\vspace{2mm}

We next investigate the modification of the
hedgehog Skyrmion profile due to further terms added to the
Skyrme energy functional. Chiral symmetry requires these additional
terms to be constructed from symmetric polynomials in the strain eigenvalues
$\lam_1^2 = f'^2$ and $\lam_2^2 = \lam_3^2 = \frac{\sin^2 f}{r^2}$.

An individual term in the energy density, before
symmetrisation, has the form
\be
K_{m,n+p} = \lam_1^m \lam_2^n \lam_3^p = f'^m \frac{\sin^{n+p} f}{r^{n+p}}
\ee
with $m$, $n$ and $p$ even integers (not all zero). Its
contribution to the energy is
\be
4\pi \int_0^\infty r^2 \left( f'^m \frac{\sin^{n+p} f}{r^{n+p}}
\right) \, dr \,,
\ee
leading to the contribution
\be
m \, \frac{d}{dr} \left( f'^{m-1} \frac{\sin^{n+p} f}{r^{n+p-2}} \right)
- \frac{n+p}{2} \, \sin 2f \, f'^m \frac{\sin^{n+p-2} f}{r^{n+p-2}}
\label{eqmn}
\ee
to the field equation for $f$. The solution for small $r$ still has
leading terms
\be
f(r) = \pi - \beta r + \gamma \beta^3 r^3 + O(r^5) \,.
\ee
Substituting this into (\ref{eqmn}), and retaining terms up to $O(r^3)$, gives
the contribution of $K_{m,n+p}$ to the field equation,
\bea
&& \beta^{m+n+p-2}\Biggl\{ -m\left( 2\beta r - 4(3m+n+p-3)\gamma\beta^3 r^3 -
\frac{2}{3}(n+p)\beta^3 r^3 \right) \nn \\
&& +(n+p)\left( \beta r - (3m+n+p-1)\gamma\beta^3 r^3 -
\frac{1}{6}(n+p+2)\beta^3 r^3 \right) \Biggr\} .
\label{solnmn}
\eea
The $O(r^5)$ part of $f$ doesn't contribute.

The corresponding symmetrised polynomial in the strain eigenvalues is
\bea
K_{m,n,p} &=& K_{m,n+p} + K_{n,p+m} + K_{p,m+n} \nn \\
&=& f'^m \frac{\sin^{n+p}f}{r^{n+p}} + f'^n \frac{\sin^{p+m}f}{r^{p+m}} + 
f'^p \frac{\sin^{m+n}f}{r^{m+n}} \,,
\eea
where $m \ge n \ge p$ to avoid overcounting, and cyclic
permulation of the indices is sufficient because $\lam_2^2 =
\lam_3^2$. Gudnason and Nitta have also worked with these
polynomials \cite{GN}. Using (\ref{solnmn}), we find that
$K_{m,n,p}$ makes the contribution
\bea
&& \Bigl\{ 10(m^2+n^2+p^2-m-n-p)\gamma + mn+np+pm \nn \\
&& \qquad \qquad - \frac{1}{3}(m^2+n^2+p^2) - \frac{2}{3}(m+n+p) \Bigr\}
\beta^{m+n+p+1}r^3 
\label{Kcontrib}
\eea
to the field equation at $O(r^3)$, and as expected, the $O(r)$ terms
have cancelled due to symmetrisation.

The Dirichlet and Skyrme energy densities are, respectively
\bea
K_{2,0,0} &=& \lam_1^2 + \lam_2^2 + \lam_3^2
= f'^2 + 2 \, \frac{\sin^2 f}{r^2} \,, \nn \\ 
K_{2,2,0} &=& \lam_1^2\lam_2^2 + \lam_2^2\lam_3^2
+ \lam_3^2\lam_1^2 = 2 f'^2 \, \frac{\sin^2 f}{r^2} + \frac{\sin^4 f}{r^4}
\,,
\eea
and according to (\ref{Kcontrib}) they contribute
\bea
&& \left( 20\gamma - \frac{8}{3} \right) \beta^3 r^3 \,, \nn \\
&& \left( 40\gamma - \frac{4}{3} \right) \beta^5 r^3
\eea
to the field equation at $O(r^3)$. Let us now supplement these
by the additional quartic energy density usually omitted
from Skyrme-type theories, with coefficient $\mu$,
\be
\mu K_{4,0,0} = \mu (\lam_1^4 + \lam_2^4 + \lam_3^4)
= \mu \left( f'^4 + 2 \, \frac{\sin^4 f}{r^4} \right)
\ee
which contributes
\be
\mu(120\gamma - 8) \beta^5 r^3
\ee
to the field equation. The combination of these energy densities
leads to the modified equation for $\gamma$,
\be
\left( 20\gamma - \frac{8}{3} + 40\gamma\beta^2 - \frac{4}{3}\beta^2
+ 120\mu\gamma\beta^2 - 8\mu\beta^2 \right) \beta^3 r^3 = 0 \,,
\label{gammabeta}
\ee
whose solution is
\be
\gamma = \frac{2}{15}
\left( \frac{1 + (\half + 3\mu)\beta^2}{1 + (2 + 6\mu)\beta^2} \right) \,,
\label{gammaquartic}
\ee
so
\be
f(r) = \pi -\beta r
+\frac{2}{15} \left(\frac{1+(\half + 3\mu)\beta^2}{1+(2 + 6\mu)\beta^2}\right)
\beta^3 r^3 + O(r^5) \,.
\label{fmod}
\ee
$\gamma$ does not have a strong dependence on $\mu$.

The large-$r$ solution for $f$, up to $O(\frac{1}{r^6})$, has the
unchanged form
\be
f(r) = \frac{C}{r^2} - \frac{1}{21} \frac{C^3}{r^6} \,.
\label{ftail'}
\ee
This is because the leading modification to the field equation from
quartic terms in the energy density like $f'^4$ and
$\frac{\sin^4 f}{r^4}$ is at $O(\frac{1}{r^8})$,
just as we found for the Skyrme term, and this only affects the solution
of the equation for the profile $f$ at $O(\frac{1}{r^8})$.

We see that the additional quartic energy density has
little effect on the Skyrmion profile expressed in terms of the
asymptotic parameters $\beta$ and $C$. Just the value of $\gamma$
changes somewhat. However, a larger change is likely
when a matching radius $R$ is determined by energy minimisation.
The matching of (\ref{fmod}) and (\ref{ftail'}) can be arranged
so that the profile and its first derivative are continuous 
at $r=R$. $R$ will depend on $\mu$, and this will in
turn affect $\beta$ and $C$. 

Similar calculations can be used to find the effect of
sextic energy densities constructed from the strain eigenvalues.
The three independent symmetric polynomials can be selected to be
\bea
K_{6,0,0} = \lam_1^6 + \lam_2^6 + \lam_3^6 &=& f'^6 + 2 \,
\frac{\sin^6 f}{r^6} \,, \nn \\ 
K_{4,2,0} = \lam_1^4 \lam_2^2 + \lam_2^4 \lam_3^2 + \lam_3^4 \lam_1^2
&=& f'^4 \frac{\sin^2 f}{r^2} + f'^2 \frac{\sin^4 f}{r^4}
+ \frac{\sin^6 f}{r^6} \,, \nn \\
K_{2,2,2} = 3\lam_1^2 \lam_2^2 \lam_3^2 &=& 3f'^2 \frac{\sin^4 f}{r^4} \,, 
\eea
and of these, $K_{2,2,2}$ is the BPS-Skyrme term. Using
(\ref{Kcontrib}) we find they contribute additional terms
to eq.(\ref{gammabeta}), respectively,
\bea
&& \left( 300\gamma - 16 \right) \beta^7 r^3 \,, \nn \\
&& \left( 140\gamma - \frac{8}{3} \right) \beta^7 r^3 \,, \nn \\
&& \left( 60\gamma + 4 \right) \beta^7 r^3 \,.
\eea
The numerator and denominator in the expression (\ref{gammaquartic})
for $\gamma$ therefore
acquire $O(\beta^4)$ corrections whose precise form depends on the
coefficients of the sextic terms in the energy functional.

In particular, the traditional Skyrme-type theories have an energy
density combining the Dirichlet, Skyrme and BPS-Skyrme
contributions. Normalising the Dirichlet and Skyrme terms as usual,
the energy density is
\be
K_{2,0,0} + K_{2,2,0} + \nu K_{2,2,2} \,.
\ee
The field equation in this case has a small-$r$ solution where
\be
\gamma = \frac{2}{15} \left(\frac{1 + \half\beta^2 - \frac{3}{2}\nu\beta^4}
  {1 + 2\beta^2 + 3\nu\beta^4}\right) \,.
\label{tradgamma}
\ee
This formula shows that $\gamma$ is quite sensitive to the value of
$\nu$ if $\beta$ is $O(1)$.

We have not calculated the effect of terms of yet higher order
(e.g. octic). These have been classified up to 12th-order by Gudnason
and Nitta \cite{GN}. However, such higher-order terms have higher
dimensions, as they involve higher powers of derivatives, and their
coefficients are suppressed by further inverse powers of the cut-off
mass. Overall, they affect the shape of the Skyrmion hedgehog profile
mainly at small $r$. Their effect is small, and certainly not
singular at short distances, and in that sense the Skyrmion is robust.

Mathematically, the parameter $C$ could be quite sensitive to the
coefficients of the EFT, but we still need to calibrate the length
scale, i.e. convert $r$ to physical length units. The strength of the
hedgehog profile's tail, for large $r$, is physically fixed by
nucleon-nucleon forces, so this tail is truly robust up to
$O(\frac{1}{r^6})$.

\section{A Universal Hedgehog Profile}
\vspace{2mm}

The simplest hedgehog profile for small $r$ is $f(r) = \pi - \beta r$.
Adding a cubic correction $\gamma\beta^3 r^3$ so that the field
equation is satisfied to $O(r^3)$ is an improvement, but $\gamma$
depends on the energy functional, so is not universal to all
chiral EFTs. It is instead desirable to have a universal profile
that includes a cubic term, and this is proposed below.

Before doing this, we recall the earlier, simpler
version, due to Ruback and the present author \cite{MRub}. Here, we
just use the leading terms for small $r$ and large $r$, and match
$f$ and $rf'$ at some radius $R$. The profile is
\bea
f(r) &=& \pi - \beta r \,, \quad r \le R \nn \\
f(r) &=& \frac{C}{r^2} \,, \quad r \ge R \,.
\eea
If we set $y = \beta R$ and $x = \frac{C}{R^2}$, then the matching
conditions become
\bea
\pi - y &=& x \,, \nn \\
y &=& 2x \,,
\eea
with solution $y = \frac{2}{3}\pi$ and $x = \frac{1}{3}\pi$. So
the profile is
\bea
f(r) &=& \pi - \frac{2\pi}{3}\frac{r}{R} \,,
\quad r \le R \nn \\
f(r) &=& \frac{\pi}{3}\frac{R^2}{r^2} \,, \quad r \ge R \,.
\label{RubNSM}
\eea
$f$ and $rf'$ are continuous at $R$, but $r^2 f''$ has the fairly
large discontinuity $2\pi$.

$R$ is not yet determined, as a spatial rescaling does not
spoil the matching conditions; it is not surprising that
two matching conditions do not uniquely determine the three parameters
$\beta$, $C$ and $R$. $R$ is determined by a global minimisation of the energy,
which requires some numerical integration \cite{MRub}. For the original 
Skyrme theory, the Dirichlet term contributes a numerical 
multiple of $R$ to the total energy, and the Skyrme term a multiple of
$\frac{1}{R}$. The energy is minimised when these contributions are
equal (a version of Derrick's theorem), and this fixes $R$. One finds
that $R = 1.26$. The estimated energy of the Skyrmion with the 
approximate profile (\ref{RubNSM}) is then $E = 151$, about $3\%$
greater than the energy for the true profile, calculated by solving
the field equation (\ref{SkhhogEq}) using a shooting method.

We now discuss our universal profile $f(r)$. The first improvement
is to include the universal $O(\frac{1}{r^6})$ term for large $r$,
as in (\ref{ftail}). The second is to include an $O(r^3)$ term for small
$r$, as in (\ref{fcubic}) or (\ref{fmod}), but for this we need to choose
the coefficient $\gamma$. Our proposal is to fix $\gamma$ so that
$f$ has a continuous second derivative at the matching radius $R$. This
approach gives the same profile for any chiral EFT, although it is less
accurate in individual cases than if $\gamma$ is fixed by solving the
field equation. The proposed profile is of the form
\bea
f(r) &=& \pi - \beta r + \gamma \beta^3 r^3 \,, \quad r \le R \nn \\
f(r) &=& \frac{C}{r^2} - \frac{1}{21} \frac{C^3}{r^6} \,, \quad r \ge R \,.
\eea

We set $y = \beta R$ and $x = \frac{C}{R^2}$ as before. Then, matching $f$,
$rf'$ and $r^2f''$ at $r=R$ leads to the constraints
\bea
\pi - y + \gamma y^3 &=& x- \frac{1}{21} x^3 \,, \nn \\
y - 3\gamma y^3 &=& 2x - \frac{2}{7} x^3 \,, \nn \\
3\gamma y^3 &=& 3x - x^3 \,.
\eea
Elimination of $y$ and $\gamma y^3$ reduces these to
\be
\pi = 5x - x^3 \,,
\label{cubicx}
\ee
and then $y = \frac{1}{7}(9\pi - 10x)$ and $\gamma y^3 =
\frac{1}{3}(\pi - 2x)$. As the local maximum value of the
right-hand side of (\ref{cubicx}) is
$\frac{10}{3}\sqrt{\frac{5}{3}} \simeq 4.30$,
this cubic equation for $x$ has three real roots. The smaller of
the positive roots gives the desired profile. One finds $x = 0.696$,
and then $y = 3.045$ and $\gamma = 0.0207$. This determines all
the coefficients except the scale $R$. The explicit form of the
profile is now
\bea
f(r) &=& \pi - 3.045 \frac{r}{R} + 0.583 \frac{r^3}{R^3} \,,
\quad r \le R \nn \\
f(r) &=& 0.696\frac{R^2}{r^2} - 0.0161 \frac{R^6}{r^6} \,, \quad r \ge R \,.
\eea
For the other two roots, either $x$ or $\gamma$ is negative, giving
the profile a wrong shape.

In the original Skyrme theory, the total energy for this universal
profile is calculated to be
\be
E = 4\pi \left(3.538R + 9.615\frac{1}{R} \right) \,,
\ee
where the first contribution is from the Dirichlet term and the second
from the Skyrme term. The minimum occurs at $R = 1.649$, giving an
energy $E = 146.6$, or equivalently $E = 1.238 \times 12\pi^2$ as
a multiple of the topological lower bound $12\pi^2$ \cite{Fad}. For this
value of $R$, $\beta = 1.847$
and $C = 1.890$. The exact profile, found using a shooting method,
has energy $E = 1.232 \times 12\pi^2$.  Although the universal profile
gives an energy accurate to $0.5\%$, $C$ is $11\%$ less than the
numerically determined value $C=2.16$. Nevertheless, this profile
is closer to the exact solution than the approximation of ref.\cite{MRub}.

For the Skyrme-type theory including the BPS-Skyrme sextic term with
coefficient $\nu$, $\gamma$ is given by the formula (\ref{tradgamma}). If
$\beta \simeq 2$ and $\nu \simeq 0.05$, then $\gamma$ is close
to its value $\gamma = 0.0207$ for the universal profile. The
universal profile should therefore work particularly well for
this Skyrme-type theory.

\section{Conclusions}
\vspace{2mm}

We have studied the asymptotic forms of the profile function $f(r)$ of a
$B=1$ hedgehog Skyrmion, for small $r$ and large $r$. A complete,
approximate profile can then be found by matching these asymptotic
profiles and their first derivatives at some radius $r=R$, with $R$
determined by energy minimisation. The analysis applies to a large range of
energy functionals of chiral EFTs, with massless pions, and not just
to the original Skyrme theory with its Dirichlet and Skyrme terms.
Chiral symmetry is imposed by requiring that the energy density is
a combination of symmetric polynomials in the strain eigenvalues of
the Skyrme field. 

The profile at large $r$ is insensitive to the details of the
energy functional; it is more sensitive at small $r$, but only at cubic
and higher order in $r$. In all cases, the Skyrmion is smooth at
$r=0$ and has no source. We have exploited these observations to
propose a universal profile for a hedgehog Skyrmion that is
continuous and has continuous first and second derivatives.

It is hoped that the ideas developed here could be used to demonstrate that
Skyrmions are relevant to the modelling of nucleons in a larger range
of chiral EFTs than the standard Skyrme-type theories, including
in those perturbative EFTs that have been exploited to calculate
nucleon-nucleon interactions and the excitation spectra of small nuclei.

\vspace{-3mm}

%%%%%%%%% Acknowledgements %%%%%%%%%%%%%
\section*{Acknowledgements}
%%%%%%%%%%%%%%%%%%%%%%%%%%%%%%%%%%%%%%%%

I'm grateful to Dionisio Bazeia and Matheus Marques for organising the
stimulating workshop ``Modern Trends in Field Theory'' in Jo\~ao Pessoa,
Brazil in March, 2024, and for the helpful comments there by Antonio
de Felice. I also thank Sven Bjarke Gudnason for comments on this paper
and for checking a calculation. This work has been partially
supported by consolidated grant ST/T000694/1 from the UK Science
and Technology Facilities Council.

%\vspace{5mm}

\end{document}